\documentclass[a4paper,revtex4]{cas-sc}

\usepackage{amssymb}
\usepackage{widetext}
\usepackage{subfig}
\usepackage{caption}
\usepackage{subcaption}
\usepackage{graphicx}%
\usepackage{multirow}%
\usepackage{amsmath,amssymb,amsfonts}%
\usepackage{amsthm}%
\usepackage{mathrsfs}%
\usepackage{xcolor}%
\usepackage{textcomp}%
\usepackage{manyfoot}%
\usepackage{booktabs}%
\usepackage{algorithm}%
\usepackage{algorithmicx}%
\usepackage{algpseudocode}%
\usepackage{listings}%
\usepackage{url}%
\usepackage{hyperref}
\usepackage{amsmath}
\usepackage{booktabs}
\usepackage{array}
\usepackage{booktabs}
\usepackage{arydshln}  
\usepackage{caption}
\usepackage[x11names]{xcolor}

\def\tsc#1{\csdef{#1}{\textsc{\lowercase{#1}}\xspace}}
\tsc{WGM}
\tsc{QE}

\begin{document}
\let\WriteBookmarks\relax
\def\floatpagepagefraction{1}
\def\textpagefraction{.001}
\shorttitle{QARA for Solving the Exact Cover Problem}
\shortauthors{X.-H. Ni et~al.}

\title[mode = title]{Quantum-Assisted Recursive Algorithm for Solving the Exact Cover Problem}


\author[label1,label2]{Xiao-Hui Ni}
\author[label1]{Jia-Cheng Fan}
\author[label1]{Ling-Xiao Li}
\author[label1]{Zi-Wen Huang}

\author[label1]{Su-Juan Qin}[orcid=0000-0002-6405-6711]
\cormark[1]
\fnmark[1]
\ead{qsujuan@bupt.edu.cn}

\affiliation[label1]{organization={State Key Laboratory of Networking and Switching Technology, Beijing University of Posts and Telecommunications},
            city={Beijing},
            postcode={100876}, 
            country={China}}

\author[label2]{Bing-Jie Xu}

\author[label2]{Wei Huang}
\cormark[2]
\fnmark[2]
\ead{huangwei096505@aliyun.com}

\affiliation[label2]{organization={National Key Laboratory of Security Communication, Institute of Southwestern Communication},
            city={Chengdu},
            postcode={610041}, 
            country={China}}

\author[label1]{Fei Gao}




\begin{abstract}
The exact cover problem is an NP-complete problem with broad applications. Studies show that although applying the Quantum Approximate Optimization Algorithm (QAOA) to this problem can yield improved solution quality with deeper circuit depth, it can limit the algorithm's applicability on noisy intermediate-scale quantum devices. To improve solution quality at shallow depth, we propose a Quantum-Assisted Recursive Algorithm (QARA) for solving the exact cover problem. QARA addresses the problem by alternately applying classical and quantum pruning. Classical pruning is a repeatable pre-processing step to simplify the problem. When the classical pruning cannot promote the problem simplification, quantum pruning is invoked. During quantum pruning, QARA extracts information from the QAOA's output state to identify the subset with the strongest selection bias. This subset is then used to prune the problem based on our problem-tailored reduction rules. Furthermore, QARA incorporates a local verification and rollback mechanism to assistively judge the effectiveness of the quantum simplification. After quantum pruning, classical pruning is applied again to the reduced problem if the remaining subsets and element set are not null. This alternating process repeats until the original problem is fully resolved. In our numerical simulations, we evaluate the performance of QARA at one-layer depth on 140 instances with subset sizes ranging from 8 to 20. Numerical results show that the probability of QARA in finding an exact solution is approximately 60\% higher than that of both QAOA and Recursive QAOA, highlighting its efficiency.
\end{abstract}


\maketitle

\section{Introduction}
The Quantum Approximate Optimization Algorithm (QAOA) \cite{QAOA} is a class of variational quantum algorithms \cite{VQA_review, vqsd, poisson, VQLS, QAS,TD-QAS} that can be implemented on the current Noisy Intermediate Scale Quantum (NISQ) devices \cite{NISQ}. Currently, QAOA has been widely applied to various combinatorial optimization problems \cite{minimum_exact_cover,minimum_vertex_cover, QLS, CVRP, UFLP}, such as Max-Cut \cite{PIL, angle_conjecture} and graph coloring \cite{XY_QAOA, MIS&graph_coloring}, and it is considered a promising algorithm for demonstrating quantum advantage \cite{quantum_advantage}. Extensive studies have shown that the performance of QAOA is strongly dependent on the layer depth \cite{maximum_k_vertex_cover, portfolio_optimization, PQA}. Specifically, QAOA tends to require more layers to enhance the solution quality \cite{INTERP, benckmark_QAOA, exact_cover1}. However, it may lead to a significant rise in the number of optimized parameters and the total circuit depth, bringing challenges in the implementation of QAOA on NISQ devices \cite{INTERP, TQA}.

\medskip
To enhance the solution quality of QAOA within shallow layer depth, some research has been proposed \cite{Adaptive_QAOA, dynamic_adaptive_qaoa, AMA-QAOA+, ma-QAOA, RQAOA, XQAOA}. One novel approach among them is the Recursive Quantum Approximate Optimization Algorithm (RQAOA), first proposed by Bravyi \textit{et al.} \cite{RQAOA}. RQAOA reduces the problem size by performing variable substitutions under the guidance of information that is extracted from the circuit's output state. This reduction process is recursively performed until the number of remaining variables falls below a predefined threshold or no cross-terms exist in the problem Hamiltonian. Then, a classical algorithm is used to solve the final reduced problem and yield a local solution. Finally, the global solution is reconstructed by backtracking from the local solution, guided by the variable substitutions applied during reduction. The core idea of quantum-information-guided recursive simplification has inspired a series of follow-up studies, which have been applied to solve various combinatorial optimization problems \cite{Reinforcement_RQAOA, QIRO, IQO, graph_coloring_RQAOA, QEG}. Notably, these studies developed problem-tailored reduction rules instead of directly adopting RQAOA's variable substitution approach for problem reduction. This is because RQAOA's general, correlation-based rules may fail to fully capture the structural properties for some problems. Researchers have found that directly applying these generic rules to problems with specific constraints often leads to infeasible solutions \cite{QIRO}.

\medskip
The exact cover problem is a well-known NP-Complete problem with constraints, and it has broad applications such as wireless network coverage \cite{wireless_network_coverage}. Vikstål \textit{et al.} \cite{exact_cover} pioneered the use of QAOA for this problem, and their work also revealed that although a deeper layer depth may improve the solution quality, it may also lead to an increase in variational parameters and circuit depths. To our knowledge, there is currently no recursive quantum algorithm specifically designed for the exact cover problem. In this paper, we propose a Quantum-Assisted Recursive Algorithm (QARA) to improve the solution quality within shallow layer depth for solving this problem. QARA addresses the problem through a novel hybrid framework that alternately applies classical and quantum pruning to recursively simplify the problem rather than only relying on the quantum simplification as done in Refs.~\cite{QIRO,IQO,QEG,RQAOA}. The details of the classical and quantum pruning are as follows.

\begin{itemize}
    \item {\textit{Classical pruning.} The classical pruning is a repeatable pre-processing step that leverages the problem's \textit{completeness} and \textit{uniqueness} constraints to reduce the problem size. By doing so, classical pruning shoulders a portion of the workload from quantum pruning and allows subsequent quantum pruning to operate on a smaller problem instance, thereby reducing the consumption of quantum resources. Notably, the classical pruning is only effective when an element covered by only one subset can be found. When classical pruning is no longer feasible, QARA invokes quantum pruning to further promote problem simplification.}
    
    \item{\textit{Quantum pruning.} During quantum simplification, in contrast to correlation-based simplification in RQAOA, QARA extracts information from the output state of QAOA to identify the subset with the strongest selection bias and its state inspired by existing recursive quantum algorithms \cite{QIRO, IQO}. This subset is then used to prune the problem based on our problem-tailored reduction rules. Specifically, if this subset prefers to be selected, all subsets sharing common elements with it are excluded, thus determining multiple subset states. Otherwise, one subset state is given. Compared with RQAOA, QARA's reduction rules can prevent multiple coverage of elements and enable the reduction of multiple variables in a quantum reduction. Furthermore, QARA incorporates a \textit{local verification and rollback mechanism} to assist in judging the effectiveness of the quantum simplification. Specifically, the rollback is triggered if the local verification fails. This quantum simplification and verification cycle repeats until the verification is passed or the maximum rollback limit is reached.}
    After quantum pruning, classical pruning is applied again to the reduced problem if the remaining subsets and element set are not null. This alternating process continues until the problem is fully resolved.
\end{itemize}

In summary, the main contributions of QARA are threefold: (1) A hybrid recursive framework that synergistically combines classical and quantum pruning; (2) Problem-tailored quantum reduction rules that prevent multiple coverage of elements and enable the reduction of multiple variables; (3) A local verification and rollback mechanism that improves solution quality.

\medskip
In the numerical simulations, the layer depth for all quantum algorithms was fixed at one. We investigated the performance of QARA on problem instances with subset set sizes ranging from 8 to 20, and compared it with the RQAOA, QAOA, and the Classical Random Recursive Algorithm (CRRA). CRRA is the classical counterpart of QARA. The key difference lies in its random decision-making when classical pruning stalls. Specifically, CRRA randomly determines the state of a subset to be one (i.e., this subset is selected). Numerical simulation results clearly reveal the superiority of QARA over these baseline algorithms. For the same problem size, QARA's probability of obtaining the exact solution is approximately 60\% higher than that of RQAOA and QAOA. In addition, the consumption of iterations for parameter optimization of QARA is approximately 81\% lower than that of RQAOA. These results highlight the effectiveness of the problem-specific reduction strategy. Furthermore, QARA's probability of obtaining the exact solution is approximately 16.4\% higher than that of CRRA, underscoring the effectiveness of quantum information (i.e., the information that is extracted from the quantum state) in guiding the problem reduction. We also conducted an ablation study to verify the effectiveness of the local verification and rollback mechanism. The results demonstrate that QARA with the rollback mechanism significantly improves both the average solution quality and success probability compared with its variant without the mechanism. This provides strong evidence for the crucial role of this mechanism in ensuring the algorithm's robustness.

\medskip
The paper is organized as follows. To help readers better understand our work, we review the definition of the exact cover problem, the design of the classical objective function, and the construction of the problem Hamiltonian when applying QAOA to solve the exact cover problem in Section~\ref{preli}. Section~\ref{QARA_intro} presents the design framework of the QARA algorithm in detail. Section~\ref{simulation} compares the performance of different algorithms on various problem instances. Finally, Section~\ref{conclusion} summarizes the main contributions of this paper.


\section{Preliminaries} \label{preli}
In this section, we review some relevant preliminary knowledge to help readers better understand our work.

\subsection{The exact cover problem}
Give a finite set of elements $U = \{ e_1, e_2, \ldots, e_n \}$, and a set $S = \{ S_1, S_2, \ldots, S_m \}$, where $S_i \subseteq U$ for $1 \le i \le m$. Each element \( e_j \in U \) is said to be covered by a subset \( S_i \in S \) if \( e_j \in S_i \). The exact cover problem asks for a subcollection \( S'' \subseteq S \) such that the subsets in \( S'' \) are pairwise disjoint and their union exactly equals \( U \). This ensures that every element in \( U \) is covered exactly once by a unique subset in \( S'' \). For instance, given \( U = \{ 1, 2, 3, 4 \} \) and \( S = \{ S_1 = \{ 1, 2 \},\ S_2 = \{ 2 \},\ S_3 = \{ 3, 4 \},\ S_4 = \{ 1, 4 \} \} \), one valid exact cover is \( \{ S_1, S_3 \} \). Together, these subsets cover all elements of \( U \) without overlap.

\medskip
To formalize the exact cover problem, let \( x_i \in \{0,1\} \) be a binary variable indicating whether subset \( S_i \in S \) is selected (\( x_i = 1 \)) or not (\( x_i = 0 \))~\cite{QAOA}. For a problem instance with \( |S| = m \), each possible subset selection corresponds to an \( m \)-bit binary string, resulting in \( 2^m \) possible configurations in total.

\medskip
For each element \( e_j \in U \) and subset \( S_i \in S \), we define a binary indicator \( c_{S_i, e_j} \), where \( c_{S_i, e_j} = 1 \) if \( e_j \in S_i \), and \( c_{S_i, e_j} = 0 \) otherwise. Once \( U \) and \( S \) are given, these coefficients are fixed. The constraint that the element $e_j$ must be covered exactly once can be expressed as
\begin{equation}
    \quad \sum_{i=1}^m c_{S_i, e_j} \, x_i = 1,
    \label{exact_cover_constraint}
\end{equation}
where $x_i \in \{0,1\}$ indicates whether subset $S_i$ is selected into $S'$.  

\medskip
Based on the exact cover constraint in Eq.~\eqref{exact_cover_constraint}, the problem can be reformulated as the minimization of the following classical objective function
\begin{equation}
    C(x) = \sum_{j=1}^n \left( \sum_{i=1}^m c_{S_i, e_j} \, x_i - 1 \right)^2,
    \label{classical_objective}
\end{equation}
where the squared terms ensure that any violation of the exact cover condition—either without-coverage or over-coverage of an element—contributes positively to the total cost, thus increasing the objective function value. The objective function \( C(x) \) quantifies how well the selection \( x = (x_1, x_2, \ldots, x_m) \) satisfies the exact cover requirement. A solution is valid if and only if \( C(x) = 0 \). The selection violates the exact cover condition if \( C(x) > 0 \).

\subsection{Quantum approximate optimization algorithm}
In the QAOA framework, the exact cover problem is mapped to the ground state (i.e., the minimal energy state) of a quantum problem Hamiltonian \( H_C \). This encoding is achieved by converting each binary variable \( x_{i} \) in Eq.~\eqref{classical_objective} to the operator \( \frac{I - \sigma_i^z}{2} \) \cite{QAOA}, where \( \sigma_i^z \) denotes acting the Pauli-Z operator on the \( i \)-th qubit. The specific formulation of the quantum problem Hamiltonian for the exact cover problem in QAOA is given by

\begin{widetext}
\begin{align}
H_C ={} & \sum_{e_j \in S} \left( \sum_{i=1}^m c_{S_i, e_j} \, \frac{I - Z_i}{2} - I \right) 
               \left( \sum_{t=1}^m c_{S_t, e_j} \, \frac{I - Z_t}{2} - I \right) \notag \\
={} & \sum_{e_j \in S} \Bigg[ 
        \left( \sum_{i=1}^m c_{S_i, e_j} \, \frac{I - Z_i}{2} \right) 
        \left( \sum_{t=1}^m c_{S_t, e_j} \, \frac{I - Z_t}{2} \right) 
        - 2 (\sum_{i=1}^m c_{S_i, e_j} \, \frac{I - Z_i}{2}) + I 
     \Bigg] \notag \\
={} & \sum_{e_j \in S} \Bigg[ 
        \sum_{i=1}^m \left( c_{S_i, e_j} \, \frac{I - Z_i}{2} \right)^2 
        + \sum_{i \ne t} c_{S_i, e_j} \, c_{S_t, e_j} \, \frac{I - Z_i}{2} \, \frac{I - Z_t}{2} - 2 (\sum_{i=1}^m c_{S_i, e_j} \, \frac{I - Z_i}{2}) + I 
     \Bigg] \notag \\
={} & \sum_{e_j \in S} \Bigg[ 
        \sum_{i=1}^m c_{S_i, e_j} \, \frac{I - Z_i}{2} 
        + \sum_{i \ne t} c_{S_i, e_j} \, c_{S_t, e_j} \, \frac{I - Z_i - Z_t + Z_i Z_t}{4} + \sum_{i=1}^m c_{S_i, e_j} \, (Z_i - I) + I 
     \Bigg] \notag \\
={} & \sum_{e_j \in S} \Bigg[ 
        \sum_{i=1}^m c_{S_i, e_j} \, \frac{Z_i - I}{2} 
        + \sum_{i \ne t} c_{S_i, e_j} \, c_{S_t, e_j} \, \frac{I - Z_i - Z_t + Z_i Z_t}{4} + I 
     \Bigg]
\label{eq:qaoa_HC}
\end{align}
\end{widetext}

\medskip
To solve the exact cover problem, QAOA begins with the ground state \( |s\rangle \) of the mixer Hamiltonian \( H_M \), and approximately evolves toward the ground state of the problem Hamiltonian \( H_C \) using a \( p \)-layer QAOA ansatz, where \( p \) denotes the layer depth. The mixer Hamiltonian is typically defined as \( H_M = \sum_{j=1}^m \sigma_j^x \), where \( \sigma_j^x \) denotes the Pauli-X operator acting on the \( j \)-th qubit. Its ground state \( |s\rangle = |+\rangle^{\otimes m} \) can be efficiently prepared by applying Hadamard gates to the initial state \( |0\rangle^{\otimes m} \).

\medskip
Each layer of the QAOA ansatz comprises two unitary operations \( \mathrm{e}^{- \mathrm{i} \gamma_i H_C} \) and \( \mathrm{e}^{- \mathrm{i} \beta_i H_M} \), where \( \gamma_i \) and \( \beta_i \) are tunable parameters for the \( i \)-th layer. The initial quantum state \( |s\rangle \) is evolved by applying a \( p \)-layer QAOA ansatz, resulting in the output quantum state
\begin{equation}
	|\psi(\boldsymbol{\gamma}_p, \boldsymbol{\beta}_p)\rangle = \left(\prod_{i=1}^p \mathrm{e}^{-\mathrm{i}\beta_i H_M} \mathrm{e}^{-\mathrm{i}\gamma_i H_C}\right) |s\rangle,
	\label{eq:output_state}
\end{equation}
where \(\boldsymbol{\gamma}_p = (\gamma_1, \gamma_2, \dots, \gamma_p)\) and \(\boldsymbol{\beta}_p = (\beta_1, \beta_2, \dots, \beta_p)\) denote the variational QAOA parameters.

\medskip
The expectation value of the output state with respect to the problem Hamiltonian \(H_C\) is defined as
\begin{equation}
	F(\boldsymbol{\gamma}_p, \boldsymbol{\beta}_p) = \langle \psi(\boldsymbol{\gamma}_p, \boldsymbol{\beta}_p) | H_C | \psi(\boldsymbol{\gamma}_p, \boldsymbol{\beta}_p) \rangle,
	\label{eq:cost_function}
\end{equation}
which can be estimated by performing repeated measurements on the output quantum state. For the exact cover problem, the ground state of the problem Hamiltonian \(H_C\) corresponds to the minimum eigenvalue, which is zero in the ideal case. Consequently, minimizing the expectation value enables QAOA to prepare a quantum state that encodes an approximate or exact solution to the problem.

\begin{center}
    \includegraphics[width=.68\textwidth]{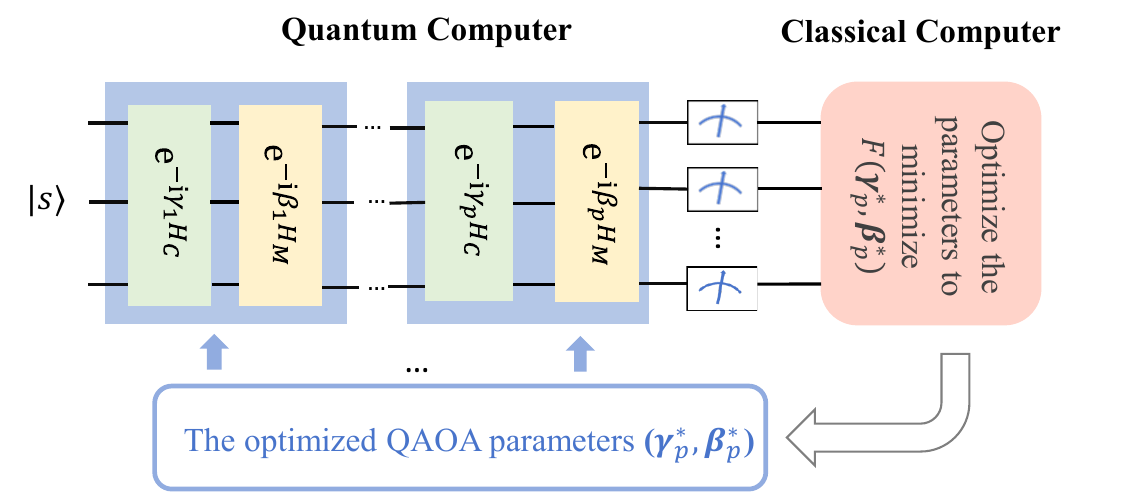} 
    \captionsetup{justification=raggedright}  
    \captionof{figure}{Schematic of a $p$-layer Quantum Approximate Optimization Algorithm. An initial quantum state  $|s\rangle=|+\rangle^{\otimes m}$ is prepared and alternately acted upon by $p$-layer QAOA ansatz. The output quantum state is measured to obtain the expectation value with respect to the $H_C$. This result is then fed to a classical optimizer, which updates the variational parameters to minimize the expectation function. This iterative quantum-classical loop continues until a predefined stopping criterion is met.} 
\label{fig:qaoa_flow} 
\end{center}

\medskip
Numerous QAOA runs are typically required to find the optimal or quasi-optimal parameters \cite{INTERP}. At the initiation of each QAOA run, the variational parameters $\gamma_i$ and $\beta_i$ for each layer $i = 1, \dots, p$ are typically initialized within the ranges $[0, 2\pi]$ and $[0, \pi]$ respectively \cite{QAOA}. The initial quantum state is then evolved by applying a series of parameterized gates. Subsequently, the quantum computer performs repeated measurements on the output state, and these outcomes are used by the classical computer to estimate the expectation value of the problem Hamiltonian. The estimated expectation value is then passed to a classical optimizer which updates the variational QAOA parameters. These updated parameters are subsequently sent back to the quantum computer to prepare the quantum state for the next round of evolution and measurement. This iterative process continues until a predefined stopping criterion is satisfied, such as the convergence of the objective function or reaching a maximum number of iterations \cite{poisson, MLI}. The procedures of QAOA are given in FIG.~\ref{fig:qaoa_flow}.                                                                             

\section{Quantum-Assisted Recursive Algorithm}
\label{QARA_intro}

QARA alternates between classical pruning and quantum pruning to tackle the original problem. Specifically, classical pruning serves as a pre-processing step, primarily leveraging the \textit{uniqueness} (i.e., each element in $U$ must only be covered once) and \textit{completeness} (i.e., each element in $U$ must be covered) properties of the exact cover problem to simplify and facilitate the problem simplification and solving. The quantum pruning is used only when classical pruning is ineffective. In the following subsection, we will introduce the process of the classical and quantum pruning in detail. Additionally, we provide a schematic diagram in FIG.~\ref{QARA} to help readers better understand the procedure of QARA, as well as the pseudocode of QARA in Algorithm~\ref{QARA_algorithm}.

\subsection{Classical pruning}

Classical pruning plays a significant pre-processing role within the QARA algorithm, and its primary purpose is to systematically reduce the scale of the original problem instance by leveraging the inherent logical properties of the exact cover problem. Specifically, the subset $S_i$ must be selected (corresponding to $x_i = 1$) if an element $e_j$ only appears in the subset $S_i$ according to the \textit{completeness} condition. According to the \textit{uniqueness} condition, any two subsets containing the same element are in conflict and cannot be selected simultaneously. If the subset $S_i$ is known to be selected, then any subset sharing common elements with $S_i$ must not be selected (corresponding to the binary variables set to $0$). Based on the above conditions of the exact cover problem, classical pruning techniques can be applied to determine some subset states before solving, thus reducing the problem size.

\medskip
Classical pruning mainly includes the process of extracting classical information and performing classical problem reduction. The details are as follows.
\begin{itemize}
    \item \textbf{\textit{Extracting Classical Information.}} In the classical pruning process, QARA first searches for such an element that is only covered by one subset by traversing the subset collection $S$, and this subset is added to a set $Q$.
    \item \textbf{\textit{Classical Problem Reduction.}} Subsequently, QARA executes classical reduction if $Q$ is not null. Specifically, according to the \textit{completeness} condition, the subset in $Q$ must be selected. Meanwhile, based on the \textit{uniqueness} condition, any subsets sharing common elements with the selected subsets must not be chosen. Following these deterministic operations, the subsets with determined states are removed from the current collection $S$, and the elements covered by the selected subset are removed from $U$, respectively. This process effectively reduces the size of the problem instance. 
\end{itemize}

In the following, the reduced problem can continue to be simplified until no further problem simplification can be achieved through classical logical deduction. The procedures of classical pruning are given in the red box in FIG.~\ref{QARA}.

\medskip
Here, an example is given to help readers understand the detailed process of the classical pruning. Give $U = \{1, 2, 3, 4\}$ and $S = \{ S_1 = \{1, 2\},\ S_2 = \{2\},\ S_3 = \{3, 4\},\ S_4 = \{1, 4\} \}$. The element $3$ appears only in the subset $S_3$. According to the completeness condition, $S_3$ must be selected. Then, by the uniqueness condition, any subset sharing elements with $S_3$, such as $S_4$, must not be selected.  After this pruning step, the remaining subsets are $S = \{ S_1 = \{1, 2\},\ S_2 = \{2\} \}$, and the universe of uncovered elements becomes $U = \{1,2\}$. For this reduced instance, classical pruning can be applied again because both the remaining subsets and uncovered elements are not null. The next round of classical pruning leads to the selection of subset $S_1$. In this example, the exact cover solution $\{S_{1},S_{3} \}$ is obtained by applying classical pruning twice, demonstrating the simplicity and effectiveness of this pruning strategy for certain instances. 


\medskip
By employing this deterministic logical inference, classical pruning significantly enhances the overall algorithm's efficiency and allows subsequent quantum pruning to operate on a smaller problem instance, thereby reducing the consumption of quantum resources. However, the effectiveness of classical pruning heavily depends on the existence of elements covered by only one subset within the given instance. Once every element appears in at least two subsets (i.e., $Q = \emptyset$), classical pruning alone cannot achieve further simplification. Therefore, additional information is required to facilitate the pruning process. 

\medskip
When classical pruning becomes ineffective, a naive approach would be to randomly assign states to certain subsets to facilitate further simplification. However, such random reductions lack consideration for the problem's structure and lead to erratic algorithmic performance. Inspired by various existing recursive quantum algorithms, we propose \textit{quantum pruning} that leverages quantum information extracted from the output state of a quantum circuit to determine the following reduced variables and their states (i.e., deciding whether a subset should be selected), thereby facilitating problem reduction. After quantum pruning, classical pruning can be invoked to further simplify the problem if the reduced problem contains elements covered by only one subset. If no such element exists, quantum pruning is applied again to drive further reduction. 

\subsection{Quantum pruning}
In each quantum pruning step, a complete QAOA run is first executed on the current problem instance to obtain the corresponding output state. Quantum information is then extracted from this output state and used to simplify the problem according to predefined reduction rules. Additionally, after each quantum simplification, we incorporate a local verification and rollback mechanism based on the properties of the problem. This mechanism is designed to assess the validity of each quantum reduction decision and improve the quality of the final solution.

\medskip
In the following, we sequentially introduce the two main components of the quantum pruning process: \textit{the quantum problem simplification} and \textit{the local verification and rollback mechanism}. The basic procedure of a single quantum pruning step is illustrated in the blue square box in FIG.~\ref{QARA}.

\subsubsection{Quantum problem simplification}


In the process of the problem simplification, two essential components are the choice of quantum information to extract from the output quantum state and how to utilize it to guide problem reduction, and these components are often problem-specific. Before presenting the specific implementation of these two components in QARA, the preparation of the circuit's output states is presented.

\medskip
\textbf{\textit{The preparation of the output state.}} In each quantum pruning step, a complete QAOA run is first executed on either the original problem or the reduced problem to obtain the output quantum state. This process consists of three steps. First, the problem Hamiltonian corresponding to the current problem instance is constructed. Second, a parameterized quantum circuit with $p$-layer QAOA ansatz is built, and its parameters are initialized. Finally, parameter optimization is performed until the expectation function converges or a predefined maximum number of iterations is reached. In our work, we choose the mixer Hamiltonian to be $H_M = \sum_{j=1}^m \sigma_j^x$, and the initial state of the circuit is $|s\rangle = |+\rangle^{\otimes m}$. Each single-layer QAOA ansatz involves only two tunable parameters, $\beta_i$ and $\gamma_i$. The initial parameters are generated randomly. The stopping criterion for the parameter optimization is that the change in the expected value of the cost function remains below $0.01$ for three consecutive iterations. This procedure corresponds to the QAOA execution illustrated in FIG.~\ref{fig:qaoa_flow}.

\medskip
Solving the exact cover problem is equivalent to determining the binary state ($0$ or $1$) of each variable $x_i$ corresponding to a subset. Inspired by various existing recursive quantum algorithms, we propose to guide problem reduction by leveraging quantum information extracted from the output state
\begin{equation}   
\vspace{-0.25cm}
|\psi(\boldsymbol{\gamma}^*_p, \boldsymbol{\beta}^*_p)\rangle = \sum_{x \in \{0,1\}^m} a_x |x\rangle
    \label{output_state}
\end{equation}
that is obtained by $p$-layer QAOA ansatz, where $a_{x}$ is the amplitude of the quantum state $|x\rangle$ and satisfies $\sum_{x \in \{0,1\}^m} |a_x|^2 = 1$.  \(\boldsymbol{\gamma}_p^* = (\gamma_1^*, \gamma_2^*, \dots, \gamma_p^*)\) and \(\boldsymbol{\beta}_p^* = (\beta_1^*, \beta_2^*, \dots, \beta_p^*)\) denote the optimized parameters.

\medskip
\textbf{\textit{Extracting quantum information from the circuit output state.}} In this paper, for each remaining subset $S_i$ in the current problem instance, we use the expectation value of the output state with respect to the Pauli-Z operator, $Z_i$, to quantify its state bias. Specifically, this expectation value is defined as
\begin{equation}
\vspace{-0.05 cm}
M_i = \left\langle \psi(\boldsymbol{\gamma}^*_p, \boldsymbol{\beta}^*_p) \middle| Z_i \middle| \psi(\boldsymbol{\gamma}^*_p, \boldsymbol{\beta}^*_p) \right\rangle = \sum_{x \in \{0,1\}^m} |a_x|^2 (-1)^{x_i}.
\end{equation}
Here, $x_i \in \{0,1\}$ represents the state of subset $S_i$ within the basis quantum state $|x\rangle$. A value of $M_i$ approaching $1$ indicates that the state of subset $S_i$ is biased toward $0$ (i.e., ``not selected''). Conversely, a value of $M_i$ approaching $-1$ suggests a bias toward $1$ (i.e., ``selected'').  After computing $M_i$ for all remaining subsets, the subset corresponding to the highest absolute bias, $S_{i^{*}} = \max_i\{|M_i|\}$, is chosen as the variable to be fixed in this step. If multiple subsets share the maximum absolute bias value, we randomly select one among them. The state of this chosen subset $S_{i^{*}}$ is then determined by the sign of $M_{i^{*}}$. Specifically, it is fixed to $0$ if $M_{i^{*}} > 0$ and to $1$ if $M_{i^{*}} < 0$.

\medskip
\textbf{\textit{Quantum Reduction.}} Let $d_i$ be the number of subsets in the current collection $S$ that share common elements with $S_{i^{*}}$. If the state of $S_{i^{*}}$ is determined to be $1$, then according to the \textit{uniqueness} condition of the exact cover problem, all $d_i$ subsets sharing common elements with $S_{i^{*}}$ must not be selected. Therefore, a single quantum reduction step can simultaneously determine the states of $1 + d_i$ subsets. Conversely, if the state of $S_{i^{*}}$ is determined to be $0$, only its state is fixed in this quantum reduction step. After determining the states of either $1$ or $1+d_i$ subsets, these subsets are removed from the current collection $S$ to form the remaining set $S'$. Additionally, if $x_{i^{*}} = 1$, the set of uncovered elements is updated as $U' = U \setminus S_{i^{*}}$, where $U$ is a collection of elements that are not currently covered. QARA thus obtains a newly simplified problem instance $P' = (S', U')$.

\subsubsection{Local verification and rollback mechanism}
The reduction rules described above ensure that no element is covered more than once. However, because the quantum information guiding the reduction is obtained from an approximate algorithm, a quantum simplification decision might incorrectly eliminate a subset critical for the final solution. This could lead to a solution that fails to cover all elements, thereby compromising the overall solution quality. To mitigate this issue, we introduce a \emph{local verification and rollback mechanism}.

\medskip
The purpose of the \emph{local verification} is to preliminarily assess the validity of the current quantum reduction. It only needs to check whether each remaining uncovered elements in $U'$ can be covered by the remaining subsets in $S'$ (i.e., verifying the \textit{completeness} of $P'$). This is a relatively efficient operation, with a time complexity of $O(|S'| \times |U'|)$. If the condition is satisfied, it simply confirms that all remaining uncovered elements can still be covered by the available subsets. However, this does not guarantee that the reduced problem $P'$ has an exact cover solution. In other words, there may still not exist a subset combination that can precisely cover each element exactly once even if every element can be covered. For example, the problem $P'$ has no vaild solution when $S' = \{ S_1 = \{1, 2\}, S_2 = \{2,3\} \}$ and $U' = \{1, 2, 3\}$.

\medskip
If an element in $U'$ has no corresponding subset in $S'$, it signals that the \textit{completeness} condition of the exact cover problem has been violated, rendering the current problem instance unsolvable (e.g., $S' = \{ S_1 = \{1, 2\} \}$ and $U' = \{1, 2, 3\}$). In such a case, it indicates that the quantum information used to guide the reduction may be misleading. To correct such erroneous reductions, QARA rejects the current reduction results and performs a new quantum reduction step. We refer to this process as the \emph{rollback operation}. The rollback operation can prevent incorrect reductions to some extent. However, excessive rollback operations will increase the runtime of a single run of quantum pruning. To mitigate this, we set a maximum number of rollback operations allowed within a single run of quantum pruning. After the final rollback operation, the algorithm accepts the results obtained by this round of quantum simplification and proceeds to the next pruning step even if the local verification still fails. We chose it instead of terminating the run, taking into account that QARA is an approximate algorithm. Terminating the run would mean no solution is found for that particular instance, essentially a failed attempt. By accepting a potentially imperfect simplification, QARA can continue its search. An imperfect step might still lead to an acceptable suboptimal solution, which is more practical than no solution at all. This design choice is a deliberate trade-off that prioritizes the algorithm's ability to yield a result over the strict guarantee of a perfect simplification.

\medskip
In summary, a key design choice of our local verification is only to check for the \textit{completeness} condition. This is because determining whether an exact solution exists is equivalent to solving the exact cover problem itself, which is NP-complete. Performing such a computationally expensive check at every step would negate the efficiency benefits of a quick, preliminary verification. Therefore, the local verification mechanism is designed as a fast sanity check to only identify and reject scenarios where a solution is definitively impossible due to a violation of the completeness condition.

\begin{center}
    \includegraphics[width=\textwidth]{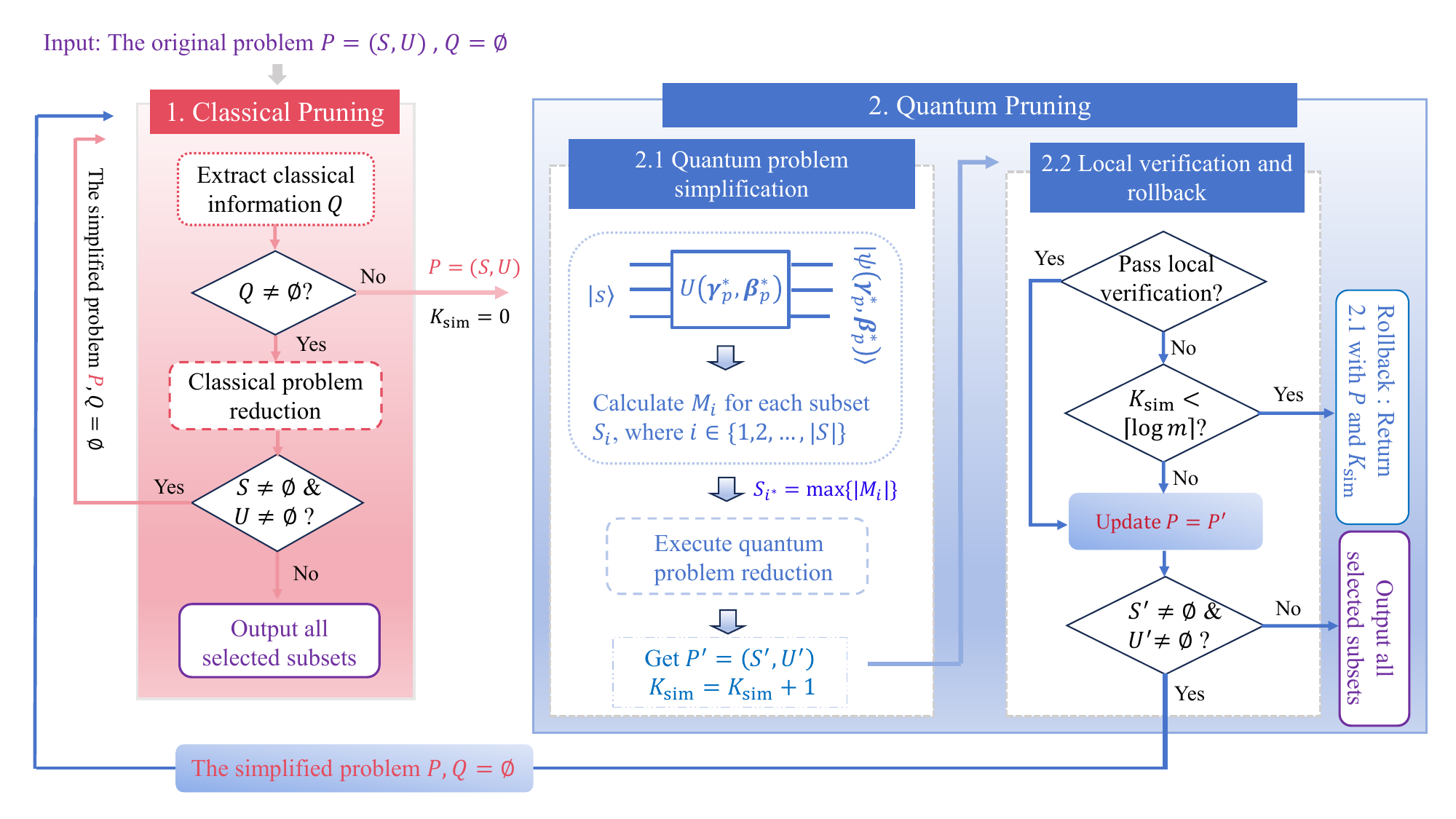} 
    \captionsetup{justification=raggedright}  
    \captionof{figure}{Flowchart of the QARA. The algorithm alternates between classical and quantum pruning to solve an exact cover problem instance. Specifically, given a problem instance $P$, the process begins with \textbf{classical pruning} as a pre-processing step. If classical pruning can no longer simplify the problem, the algorithm transitions to quantum pruning. (1) \textbf{Classical pruning phase.} In each round of classical pruning, the algorithm traverses $S$ to identify an element in $U$ that is covered by exactly one subset, and the subset is added to a set $Q$. This process is referred to as \emph{classical information extraction}. If $Q \neq \emptyset$, the problem can be simplified according to the pruning rules specified in the Algorithm~\ref{QARA_algorithm} (see Step 1). After one round of classical pruning, another round  of classical pruning can be executed to reduce the problem if the simplified problem $P = (S,U)$ satisfies that $S \neq \emptyset$ and $U \neq \emptyset$. Otherwise, QARA ends this run and outputs all selected subsets. If $Q = \emptyset$, QARA invokes quantum pruning to continue simplification. (2) \textbf{Quantum pruning phase.} This phase begins by initializing a counter $K_{\text{sim}} = 0$ to record the number of quantum simplification rounds performed. (2.1) \textbf{Quantum problem simplification.} QARA first executes one full QAOA run on the current problem $P$ to obtain the circuit output state $\left|\psi(\boldsymbol{\gamma}_p^*, \boldsymbol{\beta}_p^*)\right\rangle$. Based on this output state, quantum expectation values $M_i$ corresponding to each subset in $S$ are extracted. The subset $S_{i^*}= \max|M_{i}|$ is selected for pruning. If $M_{i^*} > 0$, set $x_{i^*} = 0$. If $M_{i^*} < 0$, set $x_{i^*} = 1$, and the states of all subsets that have the same elements as $S_{i^*}$ are simultaneously set to 0. All subsets with determined statuses are then removed from $S$, and the elements successfully covered by $S_{i^*}$ are then removed from $U$ when $x_{i^*} = 1$, resulting in a simplified pair $(S', U')$. At this point, the counter is updated as $K_{\text{sim}} \leftarrow K_{\text{sim}} + 1$. (2.2) \textbf{Local verification and rollback.} After quantum simplification, QARA checks whether the remaining elements in $U'$ can still be fully covered by the remaining subsets in $S'$. If the local verification is not passed and the upper limit on the number of rollbacks does not be achieved, the rollback operation is triggered to reconfirm the reduction of variables and their states for the current problem $P$. Otherwise, the algorithm conservatively accepts the current simplification result even though the local verification is not passed. After the quantum pruning, the algorithm re-enters the classical pruning phase if the resulting problem still contains undecided subsets or uncovered elements (i.e., $S' \neq \emptyset$ and $U' \neq \emptyset$). Otherwise, QARA ends this run and outputs all selected subsets.
 } 
    \label{QARA} 
\end{center}

\newpage
\begin{algorithm}
\caption{Quantum-Assisted Recursive Algorithm}
\label{QARA_algorithm}
\begin{algorithmic}[1]
\Statex \textbf{Input:} The initial problem $P=(S,U)$, the layer depth $p$
\Statex \textbf{Output:} All selected subsets

\State \textbf{Step 1: Classical Pruning}
\State \hspace{0.5em}  \textbf{Extract classical information}
\State \hspace{1.5em}  Initialize the set $Q = \emptyset$
\State \hspace{1.5em}  For the current problem $P=(S,U)$, seek an element that only appears in one subset $S_t$ and append this subset into $Q$, $t \in \{1,2,\dots,|S|\}$

\State \hspace{0.5em}  \textbf{If} $Q \neq \emptyset$:
\State \hspace{1.5em}  \textbf{Classical problem reduction}
\State \hspace{3 em} Set $x_t = 1$ 
\State \hspace{3 em} \textbf{If} there exist subsets $S_k$ ($k \ne t$) that share common elements with $S_t$: 
\State \hspace{3.5 em}  Set $x_k = 0$ for all such $S_k$, and then update the remaining set $S = S \setminus \{S_t, \{S_k\}\}$ and the uncovered element set $U = U \setminus S_t$

\State \hspace{3 em}  \textbf{Else}:
\State \hspace{3.5 em}  Update $S = S \setminus \{S_t\}$ and $U = U \setminus S_t$

\State \hspace{3.5em} \textbf{If} $S \ne \emptyset$ and $U \ne \emptyset$:
\State \hspace{4 em} Return \textbf{Step 1} to reduce the simplified problem $P$
\State \hspace{3.5em} \textbf{Else}:
\State \hspace{4 em} Output all selected subsets

\State \hspace{0.5em} \textbf{ Else:}
\State \hspace{1.5em} Return \textbf{Step 2} to execute Quantum Pruning

\State \textbf{Step 2: Quantum Pruning}
\State \hspace{0.5em} Initialize the number of quantum simplifications $K_{\text{sim}} = 0$ 
\State \hspace{0.5em} (2.1)\textbf{  Quantum problem simplification}
\State \hspace{1.5em} (2.1.1) Execute QAOA to solve the current problem $P = (S,U)$:
\State \hspace{2.5em} (a) Construct the problem Hamiltonian
\State \hspace{2.5em} (b) Build $p$-layer QAOA ansatz and randomly initialize the circuit parameters
\State \hspace{2.5em} (c) Execute the parameter optimization and obtain the output state $|\psi (\boldsymbol{\gamma}^*_p, \boldsymbol{\beta}^*_p)\rangle$

\State \hspace{1.5em} (2.1.2) Calculate $M_i = \langle \psi(\boldsymbol{\gamma}^*_p, \boldsymbol{\beta}^*_p) | Z_i | (\boldsymbol{\gamma}^*_p, \boldsymbol{\beta}^*_p) \rangle$ for each subset $S_{i}$, where $i \in \{ 1,2,\dots,|S|\}$

\State \hspace{1.5em} (2.1.3) Determine the subset $S_{i^*} = \max |M_i|$ and conduct quantum problem reduction 
\State \hspace{2.5em} (a) Set the state of $S_{i^*}$ to $1$ if $M_{i^*} < 0$ and $x_k = 0$ for all such $S_k$ if there exist subsets $S_k$ ($k \ne t$) that share common elements with $S_{i^*}$, and get $S' = S \setminus \{S_{i^*}, \{S_k\}\}$ and $U' = U \setminus S_{i^*}$ 
\State \hspace{2.5em} (b) Set the state of $S_{i^*}$ to $0$ if $M_{i^*} > 0$, and get $S' = S \setminus S_{i^*}$
\State \hspace{2.5em} (c) $K_{\text{sim}} \leftarrow K_{\text{sim}} + 1$

\State \hspace{0.5em} (2.2) \textbf{Local verification and rollback}
\State \hspace{1.5em} Judge each element in the current $U'$ whether it can be covered by $S'$ \Comment{Local verification}
\State \hspace{2.5em} \textbf{If} not and $K_{\text{sim}} < \lceil \log m \rceil$:
\State \hspace{3.5em} Return (2.1) \Comment{rollback}
\State \hspace{2.5em} \textbf{Else}:
\State \hspace{3.5em} Update $S = S'$, $U = U'$

\State \hspace{3.5em} \textbf{If} $S \ne \emptyset$ and $U \ne \emptyset$:
\State \hspace{4em} Return \textbf{Step 1}
\State \hspace{3.5 em} \textbf{Else}:
\State \hspace{4.0 em} Output all selected subsets
\end{algorithmic}
\end{algorithm}

\section{Numerical Results} \label{simulation}

This section presents a comprehensive comparison of algorithmic performance. We first describe the tested instances in subsection~\ref{dataset}, followed by a detailed explanation of the experimental settings in subsection~\ref{experiment settings}. The evaluation metrics used for comparison are introduced in subsection~\ref{comparison metrics}, and a thorough analysis of the numerical results is provided in subsection~\ref{analysis}. We also conducted an ablation study to verify the effectiveness of the local verification and rollback mechanism, and the results are given in subsection~\ref{ablation_study}.

\subsection{Dataset} \label{dataset}
Numerical simulations were conducted on problem instances where the number of subsets is equal to the number of elements ($m = n$). The size of the subset collection $S$ ranged from $m = 8$ to $m = 20$, with increments of two ($m \in \{8, 10, 12, 14, 16, 18, 20\}$). For each problem size, we randomly generated $20$ instances, resulting in a total of $140$ test instances. It is important to note that each instance is guaranteed to have at least one exact solution. Furthermore, in every instance, each element is required to appear in at least two distinct subsets. This condition ensures that classical pruning is ineffective at the outset, thereby forcing the algorithms (QARA, RQAOA, and CRRA) to rely on quantum or random pruning for the initial problem reduction.

\subsection{Experiment settings}\label{experiment settings}

In this work, we compare the performance of our proposed algorithm, the standard QAOA, the classical random reduction algorithm (CRRA), and RQAOA. CRRA is a classical variant of our algorithm, differing only in its pruning method. When classical pruning stalls, CRRA randomly selects a subset and sets its state to $1$ to achieve the most significant possible simplification. After all, setting the state of a certain subset to 1 may simplify multiple variables at once compared with setting it to 0. The purpose of including CRRA in the comparison is to evaluate whether quantum-information-guided reduction can offer higher accuracy and efficiency than random selection. For both QARA and CRRA, the maximum number of allowed rollbacks in a single pruning step is set to $\left \lceil \log(m) \right \rceil$, where $m$ is a variable that represents the size of the current subproblem.

\medskip
For QARA, QAOA, and RQAOA, we set the QAOA ansatz depth to $1$. Before each optimization process, the initial parameters are randomly generated. For QAOA, a single run terminates when the change in the expected value of the objective function remains below $0.01$ for three consecutive iterations. Both RQAOA and QARA may involve multiple quantum reduction steps within a single run, which results in multiple rounds of parameter optimization. In each optimization round, the choice of the initial state and the stopping criterion are the same as those used in the standard QAOA setting. For QARA and CRRA, a single run terminates when the size of the remaining set $S$ or the number of remaining uncovered elements equals zero. 

\medskip
For RQAOA, the stopping condition for a single run is either when the size of the remaining subset collection satisfies $|S| \le 5$, or when no cross terms (i.e., terms containing $Z_iZ_j$) remain in the reduced problem Hamiltonian. The latter means that further reduction based on adjacent variable correlations is no longer possible. For the remaining reduced problem, we invoke the classical random reduction algorithm to obtain an approximate solution, and then backtrack to reconstruct the global solution using the variable substitution rules recorded during the reduction process. To help readers better understand the procedure of RQAOA applied to the exact cover problem, we provide the corresponding pseudocode in Appendix~\ref{RQAOA}.

\medskip
The numerical simulations were conducted using the MindSpore Quantum 0.7.0 framework \cite{mindquantum}. We employ the Adam optimizer to train the QAOA parameters. Its ability to adapt learning rates for each parameter and incorporate momentum enables faster convergence and improved stability. 

\subsection{Comparison metrics} \label{comparison metrics}
On each problem instance, we collect the following statistics for each algorithm over $R = 50$ runs.

\medskip
The optimal classical objective value $C_{\text{opt}}$ represents the minimum objective value achieved across all runs, which is expressed as
\begin{equation}
C_{\mathrm{opt}} = \min \{ C_{1}, C_{2}, \dots, C_{R} \}.
\end{equation}
A value of $C_{\mathrm{opt}}$ closer to 0 indicates that optimal solution obtained by the algorithm is closer to the true solution of the problem.

\medskip
The average classical objective value $C_{\mathrm{avg}}$ is the mean objective value obtained across
all runs, which is expressed as
\begin{equation}
C_{\mathrm{avg}} = \frac{1}{R} \sum_{j=1}^{R} C_j.
\end{equation}
Here, $C_j$ represents the classical objective value obtained in the $j$-th run. A smaller $C_{\text{avg}}$ indicates that the algorithm consistently produces high-quality solutions across different runs, and it has a greater stability.

\medskip
The probability success $P_{\mathrm{success}}$ is the ratio of successful runs to the total number of
runs, where a run is considered successful if the exact solution (i.e., a classical objective value of 0) is obtained. This metric is given as
\begin{equation}
    P_{\mathrm{success}} = \frac{R_{\mathrm{success}}}{R}.
\end{equation}
A higher $P_{\mathrm{success}}$ implies a higher probability of obtaining the exact solution in a single run.

\medskip
The average number of iterations $T_{\mathrm{ITR}}$ is the average number of iterations consumed in a
single run of the quantum algorithm, and it is expressed as
\begin{equation}
T_{\mathrm{ITR}} = \frac{1}{R} \sum_{j=1}^{R} T_j.    
\end{equation}
Here, $T_j$ is the number of iterations for parameter optimization consumed in the $j$-th run. A smaller $T_{\mathrm{ITR}}$ indicates that the algorithm reaches the stopping criterion more quickly in each run.

\medskip
On a problem size with $K=20$ instances, we compute the mean of each performance metric to provide a robust evaluation. The mean optimal classical objective value of an algorithm is defined as
\begin{equation}
\overline{C}_{\mathrm{opt}} = \frac{1}{K} \sum_{k=1}^{K} C_{\mathrm{opt}}^{(k)}.
\end{equation}
Here, $C_{\mathrm{opt}}^{(k)}$ denotes the optimal classical objective value achieved from multiple optimization runs on the $k$-th instance. Similarly, the mean average classical objective value of an algorithm is defined as 
\begin{equation}
\overline{C}_{\mathrm{avg}} = \frac{1}{K} \sum_{k=1}^{K} C_{\mathrm{avg}}^{(k)}.
\end{equation}
A smaller $\overline{C}_{\mathrm{opt}}$ indicates that an algorithm can find higher-quality solutions in a problem size, while a smaller $\overline{C}_{\mathrm{avg}}$ suggests greater solution stability across different instances and runs.

\medskip
The mean success probability of the algorithm is defined as
\[
\bar{P}_{\mathrm{success}} = \frac{1}{K} \sum_{k=1}^{K} P_{\mathrm{success}}^{(k)}.
\]
A higher value of $\bar{P}_{\mathrm{success}}$ indicates a greater average probability of obtaining the exact solution in a single run on given problem sizes. The mean number of iterations is given by
\[
\bar{T}_{\mathrm{ITR}} = \frac{1}{K} \sum_{k=1}^{K} T_{\mathrm{ITR}}^{(k)},
\]
where a smaller value of $\bar{T}_{\mathrm{ITR}}$ implies faster convergence in the parameter optimization process on given problem sizes.

\medskip
In addition, following Ref.~\cite{QIRO}, we also report the proportion of solved instances where each algorithm successfully finds an exact solution across multiple runs. This metric is defined as
\[
S_{\mathrm{ratio}} = \frac{K_{\mathrm{exact}}}{K},
\]
where $K$ denotes the number of problem instances to be solved for each $m$, and $K_{\mathrm{exact}}$ denotes the number of instances for which the algorithm obtains the exact solution through multiple runs. The value of $S_{\mathrm{ratio}}$ lies between 0 and 1, and a larger value indicates that the algorithm can find the exact solution in more instances over $R = 50$ runs.

\medskip
For QARA, RQAOA, and CRRA, each run produces a binary string of length $m$. This string is then substituted into Eq.~\eqref{classical_objective} to compute its classical objective value. In contrast, the output of a QAOA run is a quantum superposition state. In our evaluation, we use the bitstring with the highest measurement probability from the QAOA output state as the final approximate solution and compute its corresponding objective value.

\newpage
\begin{figure}[htbp]
    \centering
    \begin{minipage}[b]{0.45\textwidth} 
        \centering
        \subfloat[The mean optimal classical objective value $\bar{C}_{\mathrm{opt}}$]{\label{mean_opt_value} \includegraphics[width=\textwidth]{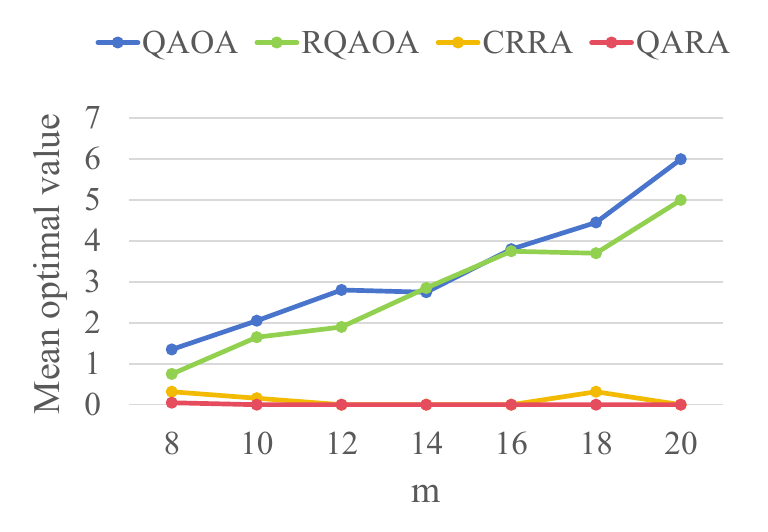}} 
    \end{minipage}
    \hspace{0.02\textwidth} 
    \begin{minipage}[b]{0.45\textwidth} 
        \centering
        \subfloat[ The mean average classical objective value $\bar{C}_{\mathrm{avg}}$]{\label{mean_avg_value} \includegraphics[width=\textwidth]{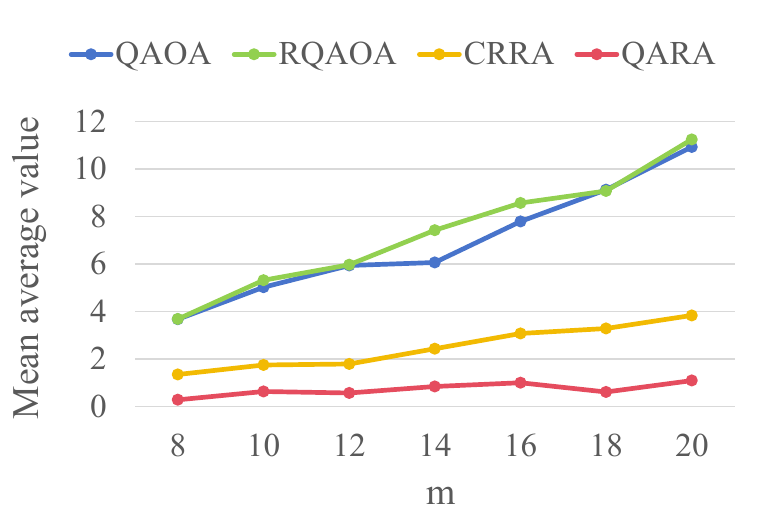}} 
    \end{minipage}
    \captionsetup{justification=raggedright}  
    \caption{Variation of $\bar{C}_{\mathrm{opt}}$ and $\bar{C}_{\mathrm{avg}}$ with problem size $m$, obtained by each algorithm over multiple runs on given instances. A smaller $\overline{C}_{\mathrm{opt}}$ indicates that an algorithm can find higher-quality solutions in a problem size, while a smaller $\overline{C}_{\mathrm{avg}}$ suggests greater solution stability across different instances and runs.
} 
    \label{mean_value} 
\end{figure}

\begin{figure}[htbp]
    \centering
    \begin{minipage}[b]{0.45\textwidth} 
        \centering
        \subfloat[The proportion of solved instances $S_{\mathrm{ratio}}$]{\label{solved_ratio} \includegraphics[width=\textwidth]{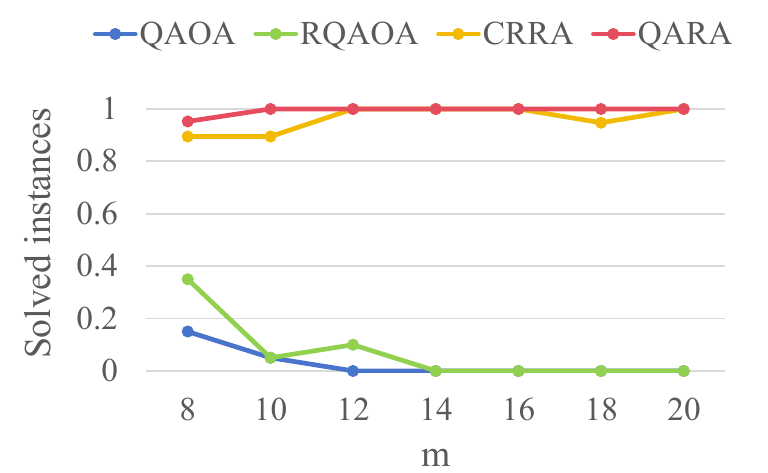}} 
    \end{minipage}
    \hspace{0.02\textwidth} 
    \begin{minipage}[b]{0.45\textwidth} 
        \centering
        \subfloat[ The mean success probability $\bar{P}_{\mathrm{success}}$]{\label{mean_success_prob} \includegraphics[width=\textwidth]{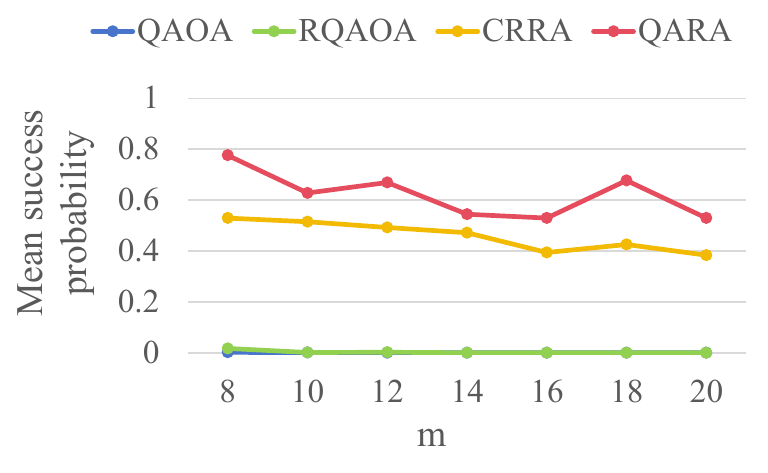}} 
    \end{minipage}
    \captionsetup{justification=raggedright}  
    \caption{Variation of $S_{\mathrm{ratio}}$ and $\bar{P}_{\mathrm{success}}$ with problem size $m$, obtained by each algorithm over multiple runs on given instances. A larger $S_{\mathrm{ratio}}$ indicates that the algorithm can find the exact solution in more instances over $R = 50$ runs. A higher value of $\bar{P}_{\mathrm{success}}$ indicates a greater average probability of obtaining the exact solution in a single run on given problem sizes.
} 
    \label{ratio_prob} 
\end{figure}

\begin{figure}[htbp]
    \centering
    \includegraphics[width=.5\textwidth]{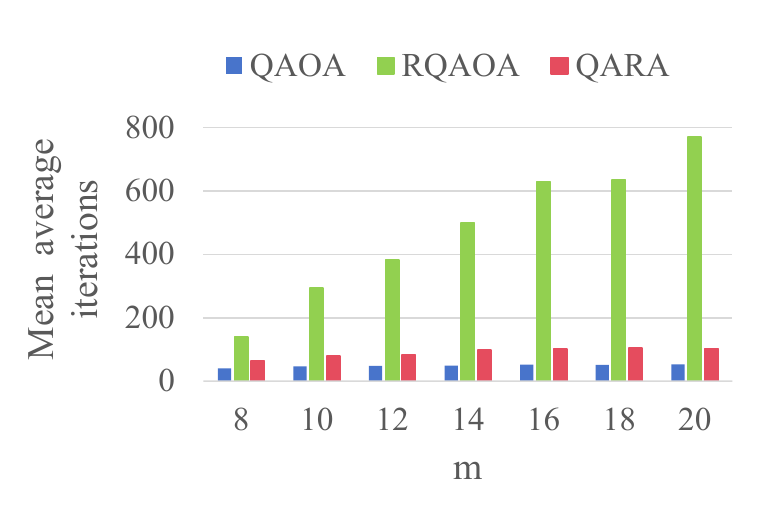} 
    
    \captionsetup{justification=raggedright}  
    \caption{Variation of $\bar{T}_{\mathrm{ITR}}$ with problem size $m$, obtained by each algorithm over multiple runs on given instances. A smaller $\bar{T}_{\mathrm{ITR}}$ implies faster convergence in the parameter optimization process on given problem sizes.
} 
    \label{mean_ITR} 
\end{figure}

\clearpage
\subsection{Performance comparison among different algorithms} \label{analysis}

FIG.~\ref{mean_value} shows the variation of $\bar{C}_{\mathrm{opt}}$ and $\bar{C}_{\mathrm{avg}}$ with problem size $m$ to evaluate the scalability of the algorithms on larger instances. The results indicate that the solution quality of QAOA and RQAOA gradually degrades as the problem size increases, which is evidenced by the rising values of $\bar{C}_{\mathrm{opt}}$ and $\bar{C}_{\mathrm{avg}}$. While RQAOA slightly outperforms QAOA, the performance gap between them is not substantial, and both are far inferior to QARA and CRRA. This indicates that methods relying solely on QAOA or RQAOA have a very limited ability to find high-quality solutions for large-scale exact cover problems. For QAOA, this is because the solution search space becomes larger, and it is more likely to fall into low-quality solution during the parameter optimization. In such cases, the bitstring with the highest probability in the QAOA output state may correspond to a low-quality approximate solution. For RQAOA, it reduces the problem size by utilizing the variable substitution. However, its reduction rules do not explicitly incorporate the specific constraints of the exact cover problem. The final global solution quality of RQAOA is affected by the accuracy of the variable substitution used in each reduction step, as well as the quality of the solution obtained by the classical algorithm. In each reduction step, the quantum local information extracted from the QAOA output state may be inaccurate. Crucially, RQAOA lacks a mechanism to mitigate such errors, which may lead the reduction process down an incorrect path. As errors accumulate through multiple reduction steps, the problem reduction deviates increasingly from the correct path, ultimately resulting in a low-quality solution.

\medskip
The numerical results in FIG.~\ref{mean_value} indicate that both QARA and CRRA consistently find exact or quasi-optimal solutions (with values very close to 0) over $R=50$ runs, even as the problem size increases. This is reflected in their $\bar{C}_{\mathrm{opt}}$ values, which remain close to zero across all instances. Their solution quality is significantly higher than that of QAOA and RQAOA, as evidenced by their substantially lower $\bar{C}_{\mathrm{opt}}$ values. Moreover, QARA exhibits higher stability than CRRA, which is evidenced by the consistently lower $\bar{C}_{\mathrm{avg}}$ values of QARA compared with CRRA. Although CRRA also leverages a problem-specific pruning strategy to find quasi-optimal solutions through multiple runs, its performance is limited by its random selection method. This random selection of subsets leads to large variations in solution quality across different runs, thereby resulting in lower stability. In contrast, QARA extracts quantum information from the QAOA output state to determine the most preferred subset and its corresponding state, which provides stronger guidance than random selection. The results further demonstrate that the performance gap between QARA and the other three algorithms becomes more pronounced as the problem size increases, highlighting its superior scalability and growing advantage.

\medskip
FIG.~\ref{solved_ratio} shows how the proportion of solved instances $S_{\mathrm{ratio}}$ varies with problem size for each algorithm. Numerical simulation results indicate that, the values of this metric for QAOA and RQAOA gradually approach 0 as the problem size increases. In contrast, the proportion of solved instances obtained by CRRA and QARA consistently remains close to 1. This suggests that under a shallow QAOA ansatz, directly using QAOA to solve the original problem often results in a highest-probability bitstring that is unlikely to be the true solution. Combined with the results in FIG.~\ref{mean_value}, this further confirms that the quality of the obtained solution is relatively poor. For small-scale instances, although RQAOA achieves a higher proportion of solved instances than QAOA, it is still significantly lower than that of QARA and CRRA. This is because the reduction rules designed in RQAOA do not incorporate the constraints of the problem, and the resulting solution may violate the \emph{uniqueness} or \emph{completeness} requirements of the exact cover problem. In contrast, the reduction strategies in QARA and CRRA are designed to always preserve the \emph{uniqueness} of the problem. To ensure \emph{completeness}, both QARA and CRRA employ a \emph{local verification and rollback mechanism}. The numerical results confirm the effectiveness of these mechanisms in significantly improving the proportion of successful exact solutions.

\medskip
FIG.~\ref{mean_success_prob} shows the variation of the mean success probability $\bar{P}_{\mathrm{success}}$ with problem size for different algorithms. For small problem sizes (e.g., $m=8$ or $10$), QAOA and RQAOA can obtain the exact solution for certain instances over multiple runs (as shown in FIG.~\ref{solved_ratio}). However, the probability of obtaining an exact solution in a single run is nearly zero. In contrast, the success probability within a single run of QARA reaches approximately 80\%. Even when the problem size increases to 20, QARA still achieves a success probability of about 58\% in a single run, which is significantly higher than that of the other three algorithms. Although both algorithms achieve a similar proportion of solved instances ($S_{\mathrm{ratio}}$) for a given problem size, QARA's single-run success probability is consistently and significantly higher than that of CRRA. This performance gap provides further evidence for the effectiveness of the quantum-information-guided reduction strategy. In summary, the quantum guidance leads to a more stable and efficient solving process.

\medskip
FIG.~\ref{mean_ITR} illustrates the variation of the mean number of iterations per run $\bar{T}_{\mathrm{ITR}}$ with the problem size $m$ for each quantum algorithm. Numerical results show that QARA consumes significantly fewer iterations per run compared withs RQAOA and slightly more than QAOA. Compared with QAOA, QARA may involve multiple rounds of parameter optimization for problem reduction within a single run, which leads to a slight increase in the total iteration count. Since RQAOA eliminates only one variable in each reduction step, a larger number of iterations is often required to reach the stopping condition (e.g., the problem size falling below a given threshold or the reduced Hamiltonian becoming non-interacting) as the problem size increases. Unlike RQAOA, QARA can eliminate multiple variables in a single reduction step, resulting in higher reduction efficiency. This substantially reduces the number of optimization rounds required, which in turn lowers the overall iteration cost.

\medskip
In summary, the numerical results collectively demonstrate the superior efficiency of QARA. Its core advantage lies in a problem-specific pruning strategy designed from the problem's constraints. Compared with the reduction method adopted by RQAOA, QARA's pruning rules can simultaneously eliminate multiple variables in a single step, which significantly reduces the number of iterations required per run. Moreover, its pruning rules ensure that the resulting solution always satisfies the \textit{uniqueness} requirement of the exact cover problem. To further enhance solution quality, QARA also incorporates a local verification and rollback mechanism. Furthermore, unlike CRRA, which also uses a problem-specific pruning strategy but makes its decisions randomly, QARA leverages quantum information extracted from a shallow QAOA quantum state to guide its pruning. This quantum-information-guided strategy significantly improves the stability and efficiency of the solving process, leading to a higher single-run success probability and a better average solution quality, thereby providing strong evidence that quantum guidance is superior to random guidance.

\subsection{Effectiveness of the local Verification and rollback Mechanism} \label{ablation_study}

In this subsection, we investigate the effectiveness of the \emph{local verification and rollback mechanism} using the aforementioned metrics on given problem instances. Since numerical results indicate that for instances of the same size, both QARA and its variant without the mechanism consistently achieve $\bar{C}_{\mathrm{opt}}$ values close to 0 and $S_{\mathrm{ratio}}$ values close to 1, we omit the corresponding visualizations. Instead, we present the results of $\bar{C}_{\mathrm{avg}}$ and $\bar{P}_{\mathrm{success}}$ for the same problem size in FIG.~\ref{mean_value_variant} and \ref{mean_success_prob_variant}.
The results demonstrate that removing the \textit{local verification and rollback mechanism} from QARA leads to a degradation in average solution quality (reflected by a higher $\bar{C}_{\mathrm{avg}}$), as well as a decrease in the probability of obtaining the exact solution in a single run. These observations clearly validate the effectiveness of the \textit{local verification and rollback mechanism}. That is, it not only improves the solution quality across different instances and runs, but also increases the likelihood of obtaining the exact solution within a single run. Although the results in FIG.~\ref{mean_ITR_variant} indicate that this mechanism may increase the iteration cost by approximately 30\% in a run, the trade-off of slightly higher iteration consumption for significantly improved algorithm's stability among various instances and runs is worthwhile.

\begin{center}
    \includegraphics[width= 0.5\textwidth]{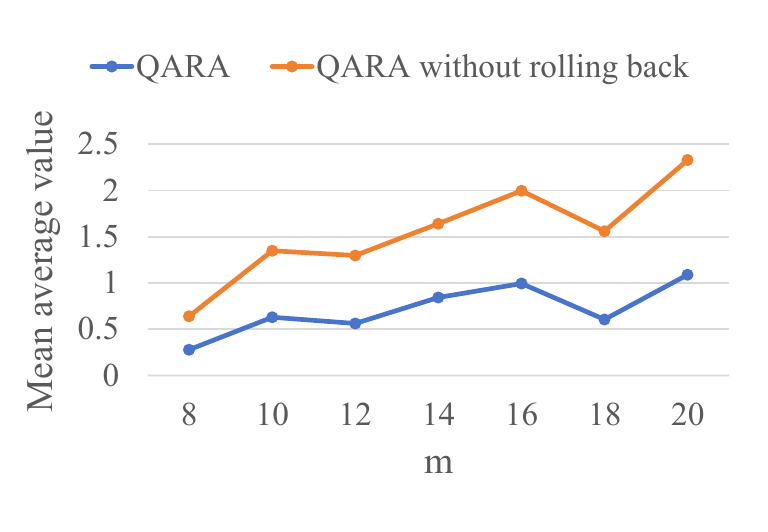} 
    \captionsetup{justification=raggedright}  
    \captionof{figure}{Variation of $\bar{C}_{\mathrm{avg}}$ with problem size $m$, obtained by each algorithm over multiple runs on given instances.
} 
    \label{mean_value_variant} 
\end{center}

\begin{center}
    \includegraphics[width= 0.5\textwidth]{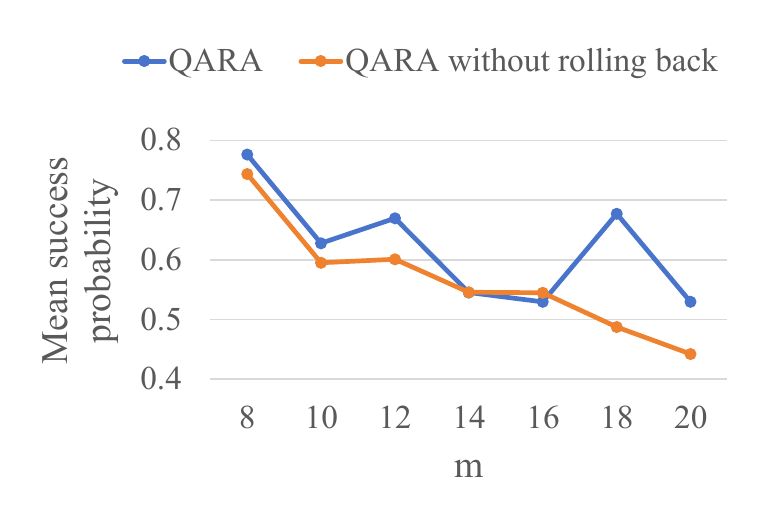} 
    \captionsetup{justification=raggedright}  
    \captionof{figure}{Variation of $\bar{P}_{\mathrm{success}}$ with problem size $m$, obtained by each algorithm over multiple runs on given instances.
} 
    \label{mean_success_prob_variant} 
\end{center}

\begin{center}
    \centering
    \includegraphics[width=0.5\textwidth]{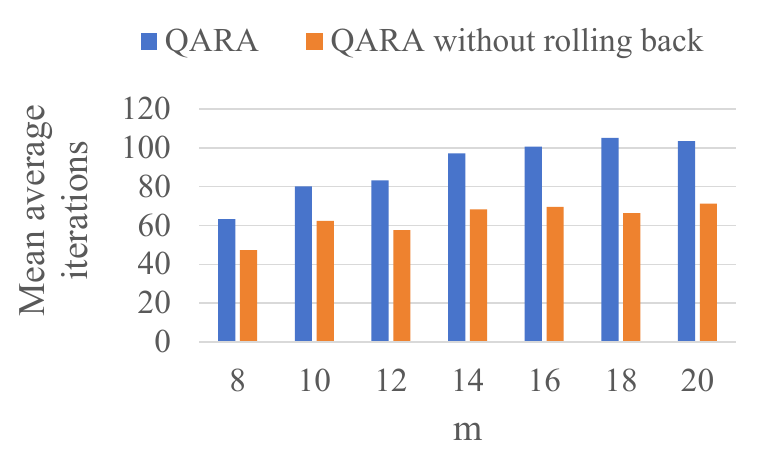} 
    \captionsetup{justification=raggedright}  
    \captionof{figure}{Variation of $\bar{T}_{\mathrm{ITR}}$ with problem size $m$, obtained by QARA and its variant over multiple runs on given instances. 
} 
    \label{mean_ITR_variant} 
\end{center}

\section{Conclusion}\label{conclusion}
This paper proposes QARA to improve the quality of solutions for the exact cover problem at shallow layer depth. The core innovation of QARA lies in its efficient recursive paradigm of alternating classical and quantum pruning. The role of classical pruning is to shoulder a portion of the workload from quantum pruning, thereby reducing the consumption of quantum resources. By leveraging efficient classical rules, it simplifies the problem whenever possible, allowing subsequent quantum pruning to operate on a smaller problem instance. This approach effectively minimizes the number of quantum runs, making the overall algorithm more resource-efficient.

\medskip
The classical pruning is a pre-processing step, and it systematically reduces the problem size by identifying uniquely covered elements. When classical pruning stalls, quantum pruning is activated. Unlike RQAOA, which reduces the problem through variable substitution, QARA extracts quantum information from the output state to determine the subset with the strongest state preference and its corresponding state, and then combines this with problem-specific pruning rules to achieve variable reduction. Furthermore, QARA creatively introduces a local verification and rollback mechanism to correct potential errors in the pruning process caused by inaccurate quantum information, further improving the correctness of the pruning path. Numerical simulation results show that QARA not only achieves efficient problem reduction but also guarantees solution quality and robustness. In conclusion, QARA provides a powerful and scalable hybrid algorithm paradigm for solving the exact cover problem using limited quantum resources on current NISQ devices.

\section{Acknowledgements}{This work is supported by the National Natural Science Foundation of China (Grant Nos. 62272056, 62372048, 62371069) and the National Key Laboratory of Security Communication Foundation (2025, 6142103042503).}

\section{Data availability}
 The data supporting this study’s findings are available from the corresponding author upon reasonable request.

\printcredits

\bibliographystyle{cas-model2-names}


\appendix
\begin{appendix}
\setcounter{figure}{0}
\setcounter{table}{0}
\setcounter{equation}{0}
\setcounter{algorithm}{0}
\renewcommand{\thefigure}{A\arabic{figure}}
\renewcommand{\thetable}{A\arabic{table}}
\renewcommand{\theequation}{A\arabic{equation}}
\renewcommand{\thealgorithm}{A\arabic{algorithm}}

\section{Recursive Quantum Approximate Optimization Algorithm} \label{RQAOA}

In this appendix, the details of solving the exact cover problem using RQAOA are provided. RQAOA is a recursive quantum-classical hybrid algorithm. Its core concept is to use quantum information to guide a recursive reduction, transforming large-scale problems into smaller ones that can be solved by classical methods. Its solution process involves three key stages, i.e., recursive simplification, classical solving, and backtracking. The pseudocode of RQAOA is given in Algorithm~\ref{alg:RQAOA}.

\begin{algorithm}
\caption{RQAOA for solving the exact cover problem}
\label{alg:RQAOA}
\begin{algorithmic}[1]

\Statex \textbf{Input:} Initial problem $P=(S, U)$, layer depth $p$, variable threshold $\theta_{\text{min}}$
\Statex \textbf{Output:} Global solution (a binary string of length $|S|$)

\State \textbf{Step 1:} Design the problem Hamiltonian $H_C$ for $P$.

\Loop
\State \textbf{Step 2:} Execute a complete QAOA run to get the output state $\left|\psi(\boldsymbol{\gamma}_p^*, \boldsymbol{\beta}_p^*)\right\rangle$

\State 2.1 Build a $p$-layer QAOA ansatz circuit based on the current problem Hamiltonian $H_C$ and the mixer Hamiltonian $H_M$, and randomly initialize the circuit parameters

\State 2.2 Execute parameter optimization to minimize the expectation value of the output state with respect to the current $H_C$.

\State \textbf{Step 3}{ Compute correlation for adjacent variables}
    \State 3.1 For any pair of cross-terms $Z_i Z_j$ in the current problem Hamiltonian $H_C$, compute the corresponding $M_{i,j} = \langle \psi(\boldsymbol{\gamma}_p^*, \boldsymbol{\beta}_p^*) | Z_iZ_j | \psi(\boldsymbol{\gamma}_p^*, \boldsymbol{\beta}_p^*) \rangle$, where $Z_i$ and $Z_j$ are called adjacent variables.

\State \textbf{Step 4:}{ Determine the most correlated adjacent variables $(Z_{i^*},Z_{j^*}) = \arg \max\{|M_{i,j}|\}$}

\State \textbf{Step 5:}{ Execute variable substitution}
    \State Replace each $Z_{i^*}$ in the current problem Hamiltonian with $\mathrm{sgn}(M_{i^*,j^*}) Z_{j^*}$, thus eliminating one variable and obtaining a new problem Hamiltonian $H_C'$.

\State \textbf{Step 6:}{ Update Hamiltonian $H_C \leftarrow H_C'$ and check stopping condition}
    \If{the number of variables in $H_C$ is less than $\theta_{\text{min}}$ or there are no cross-terms}
        \State \textbf{break} loop
    \EndIf
\EndLoop

\State \textbf{Step 7:}{ Use the CRRA method to solve the remaining reduced problem and obtain a local solution $z_{\text{local}}$}

\State \textbf{Step 8:}{ Backtrack based on $z_{\text{local}}$ and the specific variable substitution rules used during the reduction process to determine the global solution.}

\end{algorithmic}
\end{algorithm}

\medskip
\textbf{\textit{Recursive simplification.}} In each round of quantum simplification, RQAOA first builds $p$-layer QAOA ansatz according to the current problem and the mixer Hamiltonians, then it randomly initializes the QAOA parameters. During the parameter optimization, the goal is to minimize the expectation function value of the output state with respect to $H_C$. After the optimization, for any pair of cross-terms $Z_iZ_j$ in the current problem Hamiltonian, quantum information is extracted from the output state $\left|\psi(\boldsymbol{\gamma}_p^*, \boldsymbol{\beta}_p^*)\right\rangle$ by computing the correlation $M_{i,j} = \langle \psi(\boldsymbol{\gamma}_p^*, \boldsymbol{\beta}_p^*) | Z_iZ_j | \psi(\boldsymbol{\gamma}_p^*, \boldsymbol{\beta}_p^*) \rangle$. This quantum information is then used to identify the pair of adjacent variables $(Z_{i^*}, Z_{j^*})$ with the strongest correlation, defined as $(Z_{i^*},Z_{j^*}) = \max\{|M_{i,j}|\}$. Subsequently, each occurrence of $Z_{i^*}$ in the current problem Hamiltonian is then replaced with an expression involving $Z_{j^*}$. This substitution eliminates variable $Z_{i^*}$ and yields a new problem Hamiltonian $H_C$. This process of variable substitution is recursively repeated under the guidance of quantum information until the number of remaining variables falls below a predefined threshold $\theta_\text{min}$ or the current $H_C$ contains no cross-terms.

\textbf{\textit{Classical solving and backtracking.}} After the recursive simplification, a classical algorithm is employed to solve the final reduced problem $P_{\mathrm{fin}}$, which yields a local solution $Z_\text{local}$. The backtracking stage is crucial for RQAOA to ensure the acquisition of a complete global solution. During the simplification stage, the algorithm records the rules for each variable substitution. After obtaining the classical solution for the small-scale problem, RQAOA backtracks through these substitution rules. By retracing the states of the simplified variables one by one, the values of all the variables that were simplified can be determined, thereby constructing the complete global solution of the original problem.

\end{appendix}

\end{document}